\documentclass{iau_FM}
\usepackage{graphicx}

\title[JD 11.~~Pre-solar grains and AGB stars] 
{Nonadditive entropy of the gravitational wave signal from GW150914}

\author[Daniel B. de Freitas, Cleo V. da Silva \& Mackson M. F. Nepomuceno]   
{Daniel B. de Freitas, Cleo V. da Silva$^1$
 \and Mackson M. F. Nepomuceno$^2$}

\affiliation{$^1$Departamento de F\'{\i}sica, Universidade Federal do Cear\'a, Caixa Postal 6030, Campus do Pici, 60455-900 Fortaleza, Cear\'a, Brasil \\ email: {\tt danielbrito@fisica.ufc.br} \\[\affilskip]
$^2$Departamento de Ci\^encia e Tecnologia, Universidade Federal Rural do Semi-\'Arido, Campus Cara\'ubas, Rio Grande do Norte, Brazil}

\pubyear{2015}
\jname{IAUS 389 Gravitational Wave Astrophysics, Volume 1} 
\editors{David Buckley, ed.}
\begin{document}

\maketitle

\begin{abstract}
We study the first gravitational wave, GW150914, detected by advanced LIGO and constructed from the data of measurement of strain relative deformation of the fabric of spacetime. We show that the time series from the gravitational wave obeys a nonadditive entropy, and its dynamics evolve with the three associated Tsallis indices named q-triplet. This fact strongly suggests that these black hole merger systems behave in a non-extensive framework. Furthermore, our results point out that the entropic indexes obtained as a function of frequency are proper statistical parameters to determine the dominant frequency when black hole coalescence is achieved.
\end{abstract}

\firstsection 
\section{Introduction}
Gravitational waves, which are ripples in the fabric of space-time generated by the acceleration of massive objects, carry crucial information about the properties of merging black holes, such as their masses and spins (cf. \cite{AdvancedLIGO,GWOSC}). In this context, our investigation explores the hypothesis that the increase in orbital velocity as black holes merge follows a non-random pattern, suggesting that these events are not purely governed by chance. We propose that this pattern is encoded in the frequency of the emitted gravitational waves. The present study aims to test this hypothesis and provide deeper insights into the dynamics of black hole mergers and the corresponding gravitational wave emissions. To achieve this, we examine the multiscale behavior of the observed waveform deformation $h(t)$ from event GW150914, using the framework of Tsallis's non-extensive statistical mechanics (see Ref. \cite{Tsallis1}), with a particular focus on estimating the Tsallis $q$-triplet, ${q_{\rm stat}, q_{\rm sen}, q_{\rm rel}}$ (see \cite{deFreitas2,deFreitas}).

\begin{figure}
	\begin{minipage}[!]{0.45\linewidth}
\includegraphics[width=0.8\textwidth]{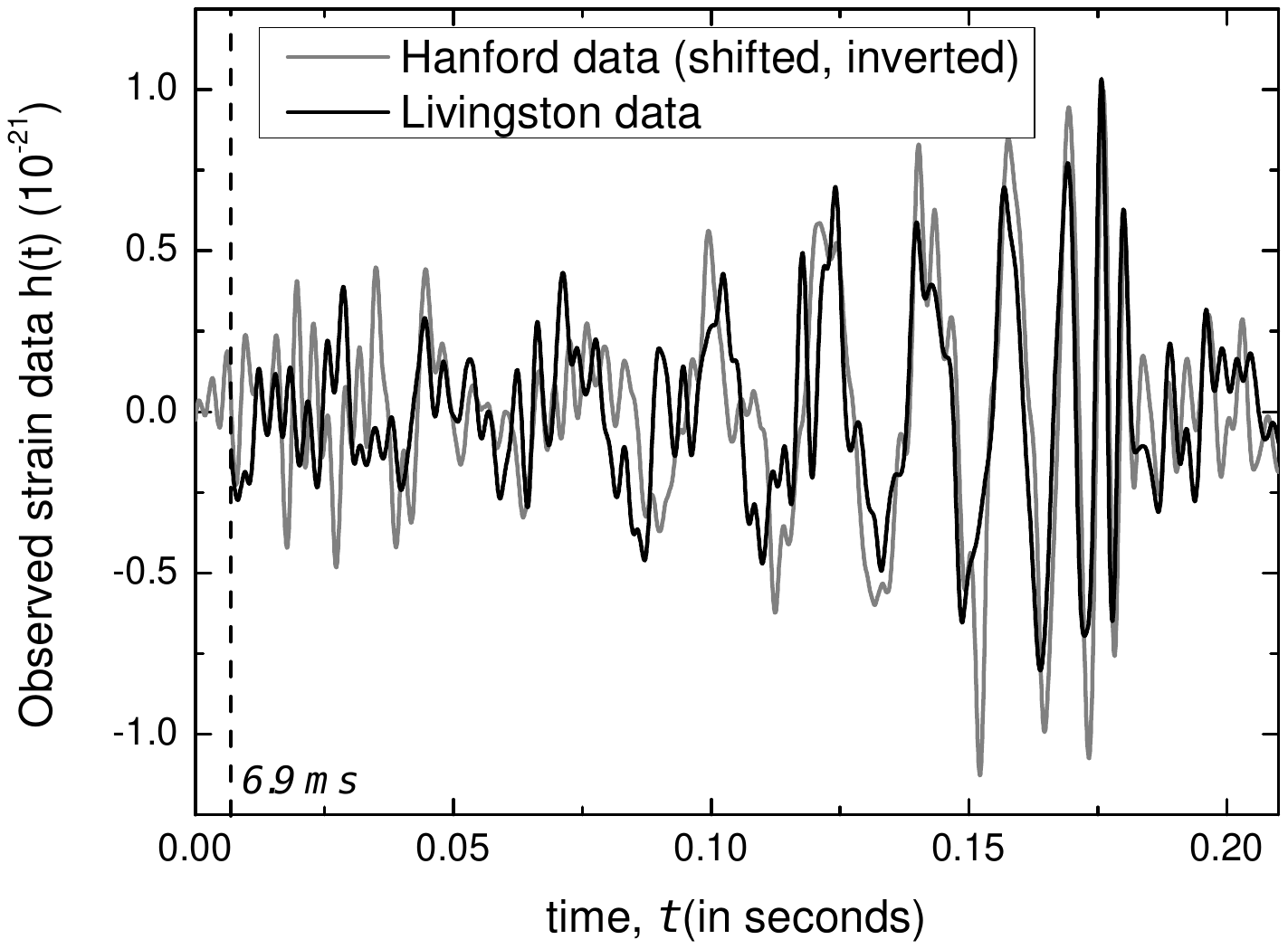}\\
\textbf{a)}
	\end{minipage}
 	\begin{minipage}[!]{0.45\linewidth}
\includegraphics[width=0.8\textwidth]{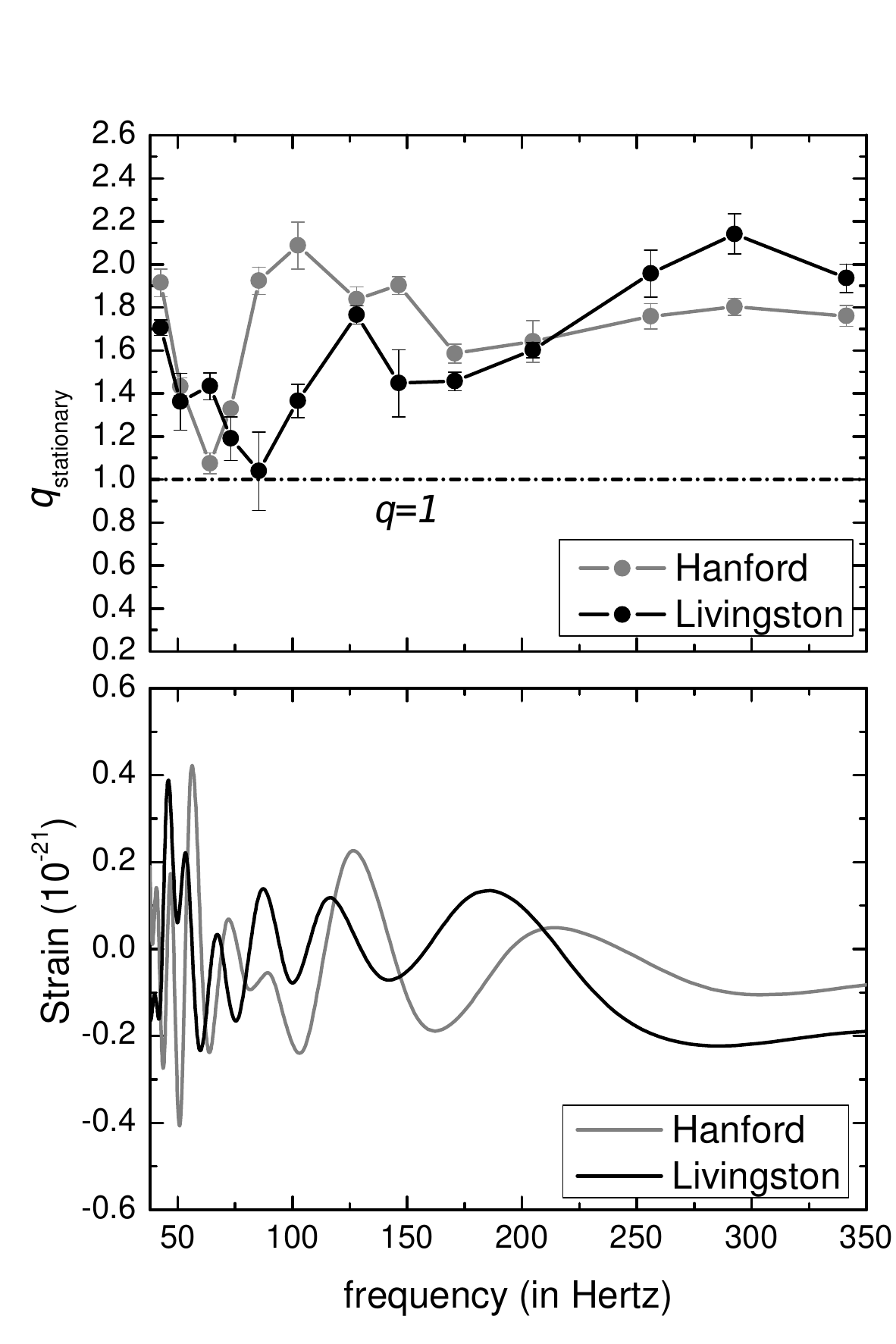}\\
\textbf{b)}
	\end{minipage}
	\caption{\textbf{a)} \textit{Left panel}: The observed strain data from the Hanford (gray) and Livingston (black) detectors, both of which have undergone bandpass and notch filtering. The Hanford strain has been time-shifted back by 6.9 ms and inverted ($-h$). The visible portion of the signal lasts approximately 0.21 seconds. The dashed line marks the 6.9 ms time shift. \textbf{b)} \textit{Top right panel}: The evolution of the entropic index $q_{stat}$ as a function of frequency for both detectors, derived from the $q$-Gaussian distribution. Here, $q=1$ corresponds to the standard Gaussian distribution. \textit{Bottom right panel}: The gravitational wave strain ($h$) as a function of frequency.} 
	\label{fig0}
\end{figure}

\section{GW150914 data and results}
The GW150914 signal consists of several cycles resembling a noisy, semi-sinusoidal time series, as depicted in Fig.\ref{fig0}a. These signals exhibit a distinct chirp, spanning from 35 to 250 Hz, with a total duration of approximately 0.2 seconds. Initially, the wave amplitude increases, starting around the 0.30-second mark. After reaching approximately 0.42 seconds, the amplitude rapidly decreases, and the frequency appears to stabilize. Accounting for a 6.9 ms time-of-flight delay at the Hanford detector in Washington (located 3000 km away from the Livingston interferometer in Louisiana), the final visible cycles indicate that the instantaneous frequency exceeds 200 Hz, as shown in Fig.\ref{fig0}a.

We implemented a novel procedure to explore the dynamical structure of the gravitational wave time series, as illustrated in Fig.~\ref{fig0}b. However, since this study is limited to analyzing a single gravitational wave event, it is necessary to extend this investigation using a more robust observational sample to address unresolved questions. Among these are discrepancies in the values of $q_{\rm stat}$ and $q_{\rm sen}$ between the detectors, particularly before and after the event, which require further exploration. These differences can only be fully assessed by applying our procedure to additional gravitational wave events.

As a result, we analyzed the time series of the GW150914 event for both the Hanford ($\rm H$) and Livingston ($\rm L$) detectors. By fitting the Tsallis distribution to observations at a frequency of 150 Hz, we obtained the following set of three indices:
$\left\{q_{\rm stat}; q_{\rm sen}; q_{\rm rel}\right\}_{\rm H}$=$\left\{1.90; 0.36; 1.40\right\}$ and
$\left\{q_{\rm stat}; q_{\rm sen}; q_{\rm rel}\right\}_{\rm L}$=$\left\{1.45; 0.63; 1.34\right\}$. These values indicate that the GW150914 gravitational wave is in an out-of-equilibrium stationary state, which $q$-statistical mechanics can accurately describe (for more details cf. \cite{dasilva}).

\end{document}